\documentclass[apj]{emulateapj}

\usepackage{epsfig}
\usepackage{amssymb}

\def\apj{{\it ApJ \,}}
\def\apjs{{\it ApJS} \,}

\def\apjl{{\it ApJ} \,}

\def\prd{{\it Phys. Rev. D.} \,}
\def\mnras{{\it MNRAS} \,}
\def\aj{{\it AJ} \,}

\def\aap{{\it A\&A} \,}

\begin{document}

\title[Dynamics of A1689]{Dynamical Study of A1689 from Wide-Field 
VLT/VIMOS Spectroscopy: Mass Profile, Concentration Parameter, and 
Velocity Anisotropy}

\author{
Doron Lemze\altaffilmark{1},
Tom Broadhurst\altaffilmark{1},
Yoel Rephaeli\altaffilmark{1,2},
Rennan Barkana\altaffilmark{1,3,4},
\& Keiichi Umetsu\altaffilmark{5,6}
}
\altaffiltext{1}
                {School of Physics and Astronomy, Tel Aviv University, Tel Aviv, 69978, Israel; doronl@wise.tau.ac.il}

\altaffiltext{2}
                {Center for Astrophysics and Space Sciences, University of California, San Diego, CA, UAS}

\altaffiltext{3}
                {Institute for Cosmic Ray Research, University of Tokyo, Kashiwa 277-8582, Japan}

\altaffiltext{4}
                {Division of Physics, Mathematics and Astronomy, California Institute of Technology, 
Mail Code 130-33, Pasadena, CA 91125, USA; Guggenheim Fellow}
\altaffiltext{5} {Institute of Astronomy and Astrophysics, Academia Sinica, P.O. Box 23-141, Taipei 106, Taiwan}
\altaffiltext{6} {Leung Center for Cosmology and Particle Astrophysics, National Taiwan University, Taiwan}

\begin{abstract}

We examine the dynamics structure of the rich
cluster A1689, combining VLT/VIMOS spectroscopy with
Subaru/Suprime-Cam imaging. The radial velocity distribution of $\sim
500$ cluster members is bounded by a pair of clearly defined velocity
caustics, with a maximum amplitude of $\sim|4000|$ km/s at $\simeq$
300 h$^{-1}$ kpc, beyond which the amplitude steadily declines,
approaching zero velocity at a limiting radius of $\sim$ 2 h$^{-1}$
Mpc. We derive the 3D velocity anisotropy and galaxy number density
profiles using a model-independent method to solve the Jeans equation,
simultaneously incorporating the observed velocity dispersion profile,
the galaxy counts from deep Subaru imaging, and our previously derived
cluster mass profile from a joint lensing and X-ray analysis. The velocity
anisotropy is found to be predominantly radial at large radius,
becoming increasingly tangential towards the center, in accord with
expectations. We also analyze the galaxy data independently of our
previous analysis using two different methods: The first
is based on a solution of the Jeans equation assuming an NFW form for
the mass distribution, whereas in the second method the caustic amplitude 
is used to determine the escape velocity. The cluster virial mass derived
by both of these dynamical methods is in good agreement with results from 
our earlier lensing and X-ray analysis. We also confirm the high NFW
concentration parameter, with results from both methods combined to
yield $c_{\rm vir}>13$ (1$\sigma$). The inferred virial radius is
consistent with the limiting radius where the caustics approach zero
velocity and where the counts of cluster members drop off, suggesting
that infall onto A1689 is currently not significant.
\end{abstract}


\section{Introduction}

Clusters of galaxies display a wide range of distinct observational
phenomena, providing detailed physical information of central
importance to cosmology. Recent observations have resulted in an
impressive consistency with the predictions of the standard
$\Lambda$CDM model (Tegmark et al.\ 2004; Spergel et al.\ 2007 and
references therein). In this model, a cosmological constant dominates
the cosmic energy budget today, but galaxies and other structures were
assembled earlier, primarily out of cold dark matter (CDM). While the
model successfully matches observations of the primary anisotropy of
the cosmic microwave background and the large-scale structure in
galaxy surveys, it is also important to test its validity on smaller
scales. The abundance and structure of non-linear objects are
sensitive probes of the properties of dark matter (DM) and of the
primordial density fluctuation field. Gas cooling and various feedback
mechanisms complicate the interpretation of the DM distribution in
galaxies (e.g., Bower et al.\ 2006). However, for clusters most of the
gas is observed to be too hot and rarefied to cool efficiently and is
thus expected to be in hydrodynamical equilibrium at the virial
temperature, and therefore to have only a small influence on the
non-baryonic mass distribution (outside the inner core). Hence it is
reasonable to suppose that constraints on the mass profile of a
cluster probe the dominant DM.

Interaction of a cluster with another cluster or group of galaxies is
commonly indicated by the presence of shock fronts, seen in high
quality X-ray observations. In the case of the ``bullet cluster'', a
cone-shaped shock front is visible indicating that two clusters have
passed through each other, with an obvious merger of the intracluster
(IC) gas of the two clusters, but the galaxies and the lensing mass
distribution are largely intact. This directly implies that the DM is
collisionless like the galaxies (Markevitch et al.\ 2002; Clowe et al.\
2004; Brada{\v c} et al.\ 2006), supporting the simplest possibility
that DM interacts only via gravity.

For relaxed clusters, knowledge of the mass profile can provide
crucial information regarding the nature of DM and the thermal history
of IC gas. Precise measurements have been made by dynamical analysis
of galaxy velocities, via hydrostatic analysis of X-ray observations,
or directly via lensing. Detailed measurements of the X-ray spectrum
and intensity profile yield the distribution of total mass through
solving the equation of hydrostatic equilibrium. In practice the
temperature measurements are often made uncertain by significant
complications such as the likely multiphase nature of gas in
equilibrium. We have recently developed a model-independent joint
lensing/X-ray analysis (Lemze et al.\ 2008, hereafter L08) to examine
the consistency of the X-ray temperature and surface brightness
profiles with the lensing data, finding that the cluster's spectrally
measured temperature profile is systematically $\sim 30-40\%$ lower
than deduced from solving the equation of hydrostatic equilibrium with
the precisely measured lensing mass profile. This may reflect in part
the ambiguity in deriving 3D temperatures from projected X-ray data,
since a given line of sight will in general intersect a range of gas
temperatures if the gas is not strictly isothermal (Mazzotta et al.\
2004; Vikhlinin 2006). It is also conceivable that the gas is not
strictly single phase but may contain small scale structure, including
relatively dense cooler clouds as found in detailed simulations
(Kawahara et al.\ 2007), which may lead to a significant downward bias
in observed temperature estimates.

The velocity dispersion profile of clusters has long been used to
estimate the galaxy dynamics and the cluster mass profile via the
Jeans equation, when sufficient redshift information is
available. This technique was originally applied to several well
studied nearby clusters (Fuchs \& Materne 1982; Sharples et al.\
1988), and more recently to the results of dedicated surveys (Carlberg
et al.\ 1997; Katgert et al.\ 2004; Biviano \& Katgert 2004; Hwang \&
Lee 2008). In these studies the expected velocity anisotropy
complicates the interpretation, as orbits are not expected to be
isotropic but to become predominantly radial towards the cluster's
virial radius. More recently, with the availability of larger samples
of redshift measurements, it has been recognized that relaxed clusters
should have sufficiently well defined velocity "caustics", providing
an independent means of deriving cluster mass profiles; this has been
applied to several clusters and compared with lensing-based masses
(Geller, Diaferio, \& Kurtz 1999; Rines et al.\ 2003; Diaferio,
Geller,\& Rines 2005). Joint studies of galaxy dynamics and lensing in
the central cluster region are also providing detailed mappings of the
central mass distributions of massive clusters, where the velocity
dispersion profile of the central galaxy can be compared with the
analysis of multiply imaged sources (Sand et al.\ 2008).

Lensing work is now able to measure cluster mass profiles with
sufficient precision to usefully test the distinctive prediction of a
relatively shallow mass profile for halos dominated by CDM (Navarro,
Frenk, \& White 1997, hereafter NFW; Hennawi et al.\ 2007; Duffy et
al.\ 2008). Combined weak and strong lensing measurements have shown
that the continuously steepening form of the NFW profile is a
reasonable description for the mass profiles of three carefully
studied clusters (Kneib et al.\ 2003; Gavazzi et al.\ 2003; Broadhurst
et al.\ 2005a, hereafter B05a), although with surprisingly high values
derived for the profile's characteristic concentration parameter in
each case (B05a). More recently, the concordance $\Lambda$CDM model
has been examined with increasing (though still relatively small)
samples of clusters, indicating that the concentrations derived are
significantly higher than predicted over a wide range of cluster mass,
after a statistical correction for lensing-induced biases. This is
seen both in terms of the size of the Einstein radius (Broadhurst \&
Barkana 2008) and the weak lensing profiles of several well known
clusters (Broadhurst et al.\ 2008). This question has also been
examined recently by stacking the lensing signal of cluster samples
identified in the SDSS survey. A wide dispersion in derived
concentrations is found, with an average value in better agreement
with the predicted relation (Mandelbaum et al.\ 2008). Here it is
crucial that reliance on photometric redshifts does not result in the
misclassification of cluster members as background galaxies, which
would artificially decrease the central lensing signal thereby
lowering measured concentrations.

Here we concentrate on the internal dynamics of A1689, one of the
best-studied clusters, using a relatively large redshift survey from
the VLT/VIMOS wide-field instrument. This cluster is attractive as it
is very massive and appears to be relaxed, with a highly symmetric X-ray 
morphology (L08; Riemer-Sorensen et al.\ 2008, see their fig. 1), the 
centroid of which coincides both with the cD galaxy and 
the center of mass derived from lensing (Xue \& Wu 2002; L08). Only a low 
level of substructure is visible in detailed lensing and X-ray maps 
(B05a; Broadhurst et al.\ 2005b; L08; Umetsu \& Broadhurst 2008), though Andersson \& 
Madejski (2004) claim an indication for a small deviation from a relaxed 
state. Earlier dynamical analyses were made by den Hartog \& Katgert\ 
(1996), Czoske (2004), and Lokas et al.\ (2006). In our earlier work we found 
that the cluster has a relatively high concentration parameter, which makes this 
a very interesting target to explore with independent dynamical means. 
We also make use of the extensive VLT/VIMOS data to explore galaxy dynamical 
properties in unprecedented detail, including in particular the velocity anisotropy 
profile, which has been difficult to determine from previous observations 
(Benatov et al.\ 2006).

This paper is organized as follows. In \textsection~\ref{The data
sets} we describe the data, from which we derive the radial profile of
the projected surface density (\textsection~\ref{Galaxy surface number
density}), and the projected velocity distribution and radial profile
of the velocity caustics (\textsection~\ref{Projected Velocity
Dispersion}). In \textsection~\ref{Methodology} we detail our method
of using the Jeans equation together with previous knowledge of the
cluster mass profile to determine the three-dimensional radial
profiles of galaxy velocity dispersion and anisotropy, with results
given in \textsection~\ref{The galaxy dynamical properties}. In
\textsection~\ref{Deriving the mass profile} we independently
determine the cluster mass profile from the data on galaxy dynamics,
using the caustics as a measure of the escape velocity from the
cluster (\textsection~\ref{Mass profile from velocity caustics}),
or fitting an NFW profile to the dynamical data
(\textsection~\ref{mass profile from the Jeans equation}).  We
compare in \textsection~\ref{The edge of the cluster} various
definitions of the cluster boundary, based on the radial profiles of
the cluster mass, galaxy density, or caustics, and end with a discussion in 
\textsection~\ref{Discussion}.

\section{Data Analysis}
\label{The data sets}

We use several different types of data in this paper: galaxy positions
and velocities, gravitational lensing, both strong and weak, and X-ray
emission and spectroscopy. Previously we have made joint use of the
same lensing and X-ray data to derive a cluster mass profile by a
model-independent joint analysis as described in L08. Here we extend
this work and analyze independent dynamical information on the galaxy
velocities and also the number density profile of cluster members. 
In this section we explain in detail our analysis of the latter two data sets.

\subsection{Galaxy surface number density}
\label{Galaxy surface number density}

Establishing the form of the projected profile of cluster member
galaxies is essential for fully employing the Jeans equation (Binney
\& Tremaine 1987; see \textsection~\ref{Methodology}), which allows for a
spatial distribution of galaxies that need not follow the dominant
DM. Measurement of the projected galaxy distribution requires
subtraction of the background and foreground field galaxy populations.
This must be achieved with accurate multi-color photometry as
spectroscopy is usually limited to small samples and does not extend
faint enough to include the majority of cluster members.

We used Subaru photometry in the V and I bands. The data 
are complete to a depth of $I_{AB}=26.5$, reaching eight magnitudes below 
$L^{*}$ of the luminosity function. We selected a color region based 
on a color-magnitude relation analysis, which comprises 17474
galaxies. This color region ranges from the red side of the E/SO
cluster sequence to a blue boundary chosen so that the sample extends
sufficiently blueward to include most ($\sim 85\%$) of the cluster
members as described in Medezinski et al.\ (2007, hereafter M07,
fig.~1). M07 established that this color region contains the majority
of cluster members in addition to some background galaxies, by
examining the weak lensing signal. In this color region, which
includes the E/SO sequence and bluer objects (including some
background galaxies) the lensing signal was found to be significantly
lower than the true background signal as measured for red background
galaxies. This is because (unlensed) cluster members dilute the weak
lensing signal of the background, allowing us to identify the region
of color space occupied by cluster members.

We derived the projected galaxy distribution of cluster members from
the above color-selected galaxy sample based on fitting the radial
distribution. Galaxies were first radially binned into annuli and the 
surface number density in each annulus was determined; note that at large 
radii only part of each annulus was covered by the detector. Assuming 
Poissonian ($\sqrt{N}$) errors, the galaxy surface number density was modelled 
as that due to the cluster plus a background galaxy surface number density, 
with the latter assumed uniform. Thus, we effectively neglected fluctuations 
in the background galaxy surface number density on scales smaller the cluster size (about 
$22^\prime$ at $z=0.183$). There could be such fluctuations at some level 
due to correlated structures (such as filaments) along the line of sight 
to the cluster. Galaxies in these structures could mistakenly be included 
as cluster members, a possibility that we assess below.

\begin{figure}[h]
\centering
\epsfig{file=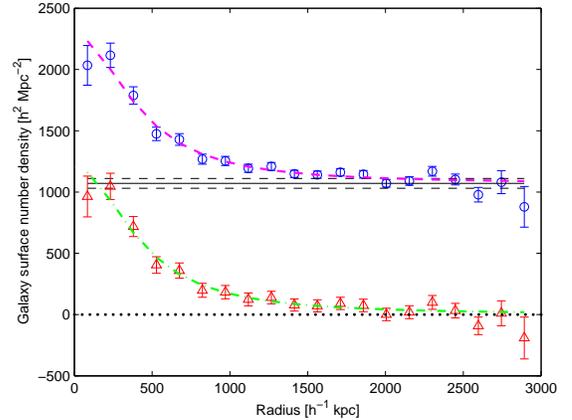, width=8cm}
\caption{Galaxy surface number density. We show the total (blue
circles), the cluster contribution (red triangles), and the background
value (solid horizontal line) and its uncertainty (dashed lines). All
uncertainties are 1-$\sigma$. We also show the core profile that best
fits the cluster galaxy surface number density (dash-dotted curve), and
the corresponding fit to the total galaxy surface number density
(dashed curve).
\label{galaxy surface number density}}
\end{figure}

Figure~\ref{galaxy surface number density} shows that the radial
profile of the above color-selected galaxy sample is well fitted with
a uniform background plus a general cored profile to represent
the cluster galaxy surface density profile:
\begin{equation}
\Sigma_{\rm tot} = \Sigma_{\rm gal} +C_{\rm bg} = \frac {\Sigma_0} {[1+(r/r_c)^2]^p} +C_{\rm bg}\ , 
\label{Sigma_tot}
\end{equation}
where $\Sigma_{\rm tot}$ is the total galaxy surface number density,
$\Sigma_0$, $r_c$, and $p$ are the three parameters of the cored profile, 
and $C_{\rm bg}$ is the background density. The resulting fit is good, 
$\chi^2/{\rm dof}=19.2/16$. Using a larger number of bins
than $N=20$ improves the reduced $\chi^2$, e.g., $N=50$ and $N=100$
give $\chi^2/{\rm dof}=36/46$ and $\chi^2/{\rm dof}=96/96$,
respectively. The level of the background is stable and does not
depend on the number of bins, e.g., for $N=20$, 50, and 100 we find
$C=1071\pm 40$, $1064 \pm 42$, and $1063\pm 41$, respectively. We
settled on just 20 bins, since this made it easier to combine 
the surface density 
data set with the projected velocity dispersion data set, which is 
inferior in terms of signal to noise per radial bin (see below). With
20 bins, we found best-fit values of $\Sigma_0=1200 \pm 140$ h$^2$
Mpc$^{-2}$, $r_c = 450 \pm 150$ h$^{-1}$ kpc, and $p=0.74 \pm 0.30$.

\subsection{Projected Velocity Dispersion}
\label{Projected Velocity Dispersion}
\label{sec:vel}

To measure the mass profile of galaxy clusters using galaxy motions it
is necessary to obtain precise velocity measurements for a
statistically large sample of galaxies. The data used here are part of
an extensive multi-object spectroscopy survey carried out with the
VIMOS spectrograph on the VLT (Czoske 2004); for observational details,
see Czoske (2004). This dataset constitutes 1469 objects with reliable
spectroscopic redshifts, a major advance over previous surveys 
of A1689. Note though that we did not try to find the galaxy surface
density profile from the projected velocity data set, since the data
set used in the previous section contains a much larger number of
galaxies.

We analyzed the spectroscopic sample by first defining cluster
membership using the velocity ``caustics'', which are clearly 
visible for this cluster in the form of a boundary which varies with radius,
peaking at around $\pm 4000$ km/s at $\sim 300$ h$^{-1}$ kpc and
declining steadily at larger radius (see figure~\ref{velocity space
diagram}). The caustics are related to the escape velocity from the
cluster and thus provide a tangible physical basis by which we can 
separate cluster members from foreground and background galaxies. 
Defining membership is especially important for massive clusters like 
A1689, since they have a relatively wide internal velocity spread and 
they extend to large radii; this increases the chance for interlopers, 
which can have a large effect on the derived projected velocity dispersion, 
especially at large radii where the number density of cluster members is low. 
Wojtak \& Lokas (2007) showed that without the proper removal of 
interlopers the inferred parameters of the mass distribution in the 
cluster are strongly biased towards higher mass and lower concentration.
See also Wojtak et al.\ (2007) for interesting comparison between different 
methods of interlopers removal, which unfortunately do not include the D99 
caustics approach used in this paper.

To define the caustics we employed the technique pioneered by Diaferio
(1999, hereafter D99) based on a multidimensional adaptive kernel
method (Silverman 1986; Pisani 1993; Pisani 1996). A summary of the
application of this technique can be found in Diaferio, Geller, \&
Rines (2005); it has previously been applied to several clusters
(Reisenegger et al.\ 2000; Biviano \& Girardi 2003; Diaferio, Geller,
\& Rines 2005; Rines \& Diaferio 2006). Briefly, a redshift-space
diagram is constructed, i.e., the line of sight velocity $v$ is
plotted versus the projected distance $R$ from the cluster center, and
for a well defined cluster these points should be distributed in a
characteristic ``trumpet'' shape, the boundaries of which are termed
caustics (Kaiser 1987; Regos \& Geller 1989). The D99 procedure
locates the caustics and determines the radial dependence of their
amplitude (in units of velocity), which is related to the escape
velocity and thus depends on the mass profile. Galaxies that are
inside the caustics can then be considered to be cluster members. 
Of course, some of these galaxies might be interlopers, but their number 
is typically a few percent and has little effect on dynamical analyses 
(A. Diaferio, private communication).

To apply this procedure we determined the threshold $\kappa$ that
defines the caustic location through $f_q(R,v)=\kappa$ (D99). Here
$f_q(R,v)$ is the galaxy density distribution in the redshift-space
diagram, smoothed with an adaptive kernel, and $v$ is the peculiar
velocity (measured with respect to the cluster's mean redshift). The
parameter $q$ sets the scaling between the quantities $R$ and $v$
within the smoothing procedure. We used $q=27$, close to the value
usually used, $25$ (D99; Rines et al.\ 2003). We note that
different values of $q$ in the $10-50$ range have little effect on the
results (D99).

The parameter $\kappa$ was chosen by minimizing the quantity
$S(\kappa,\langle R\rangle)=|\langle v_{esc}^2\rangle_{\kappa,\langle
R\rangle}-4 \langle v^2\rangle|^2$, where $\langle
v_{esc}^2\rangle_{\kappa,\langle R\rangle}= \int_0^{\langle
R\rangle}A^2(R)\phi(R)dR/\int_0^{\langle R\rangle}\phi(R)dR$ is the
mean value of the square of the caustic amplitude $A(R)$ within
$\langle R\rangle$ (the mean projected radius of the cluster members),
$\phi(R)=\int f_q(R,v)dv$, and $\langle v^2\rangle$ is the
one-dimensional velocity dispersion of the cluster members. The
uncertainty of $A$ is proportional to the inverse of the galaxy number
density within the caustics, and was estimated as $\delta A(R)/A(R) =
\kappa/ \max \left\{f_q(R,v)\right\}$, where the maximum is found
along the $v$-axis at each $R$ (D99). 

\begin{figure}[h]
\centering
\epsfig{file=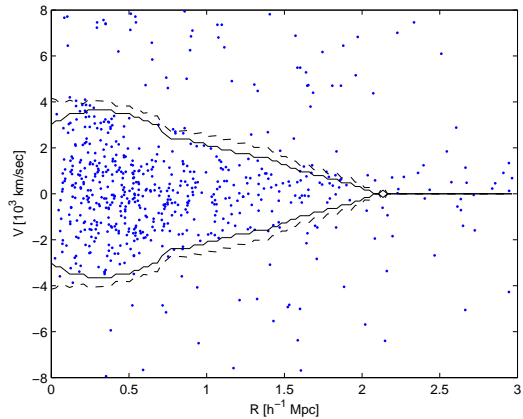, width=8cm}
\caption{Velocity-space diagram of A1689. The caustics are shown
along with 1-$\sigma$ error bars (shown only on the outer side of the
caustics for clarity). \label{velocity space diagram}}
\end{figure}

Figure~\ref{velocity space diagram} shows the velocity-space diagram
and radial profile of the derived velocity caustics. There are a total
of $476_{-43}^{+27}$ galaxies identified as cluster members lying between 
these caustics. The velocity-space diagram clearly looks qualitatively 
different inside and outside the caustics, indicating that the caustic-finding 
procedure has identified a physically-meaningful boundary. The density of 
the galaxies that lie between the caustics shows a smooth decline with 
distance from the center, indicating that the cluster has a simple monolithic 
dynamical structure. The radius at which the caustics meet at zero velocity 
represents an effective maximum radius for the cluster, which physically includes 
cluster galaxies both within the virial radius and others that may be
currently infalling onto the cluster (see the discussion in
\textsection~
\ref{Discussion}). Note that the two caustics are by definition symmetric 
about $v=0$ in this method, which figure~\ref{velocity space diagram}
clearly shows is a valid assumption for this cluster .

\begin{figure}[h]
\centering
\epsfig{file=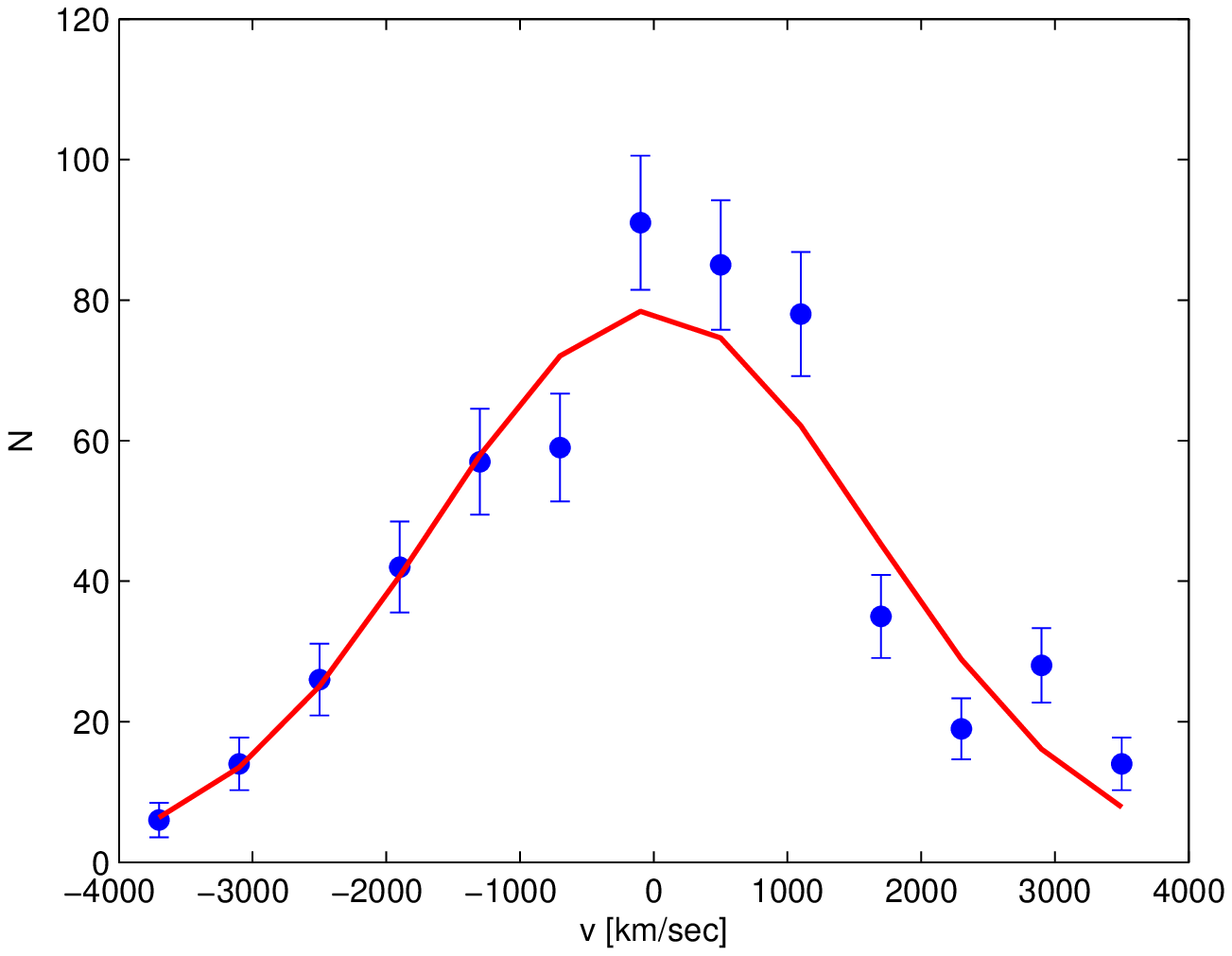, width=8cm}
\epsfig{file=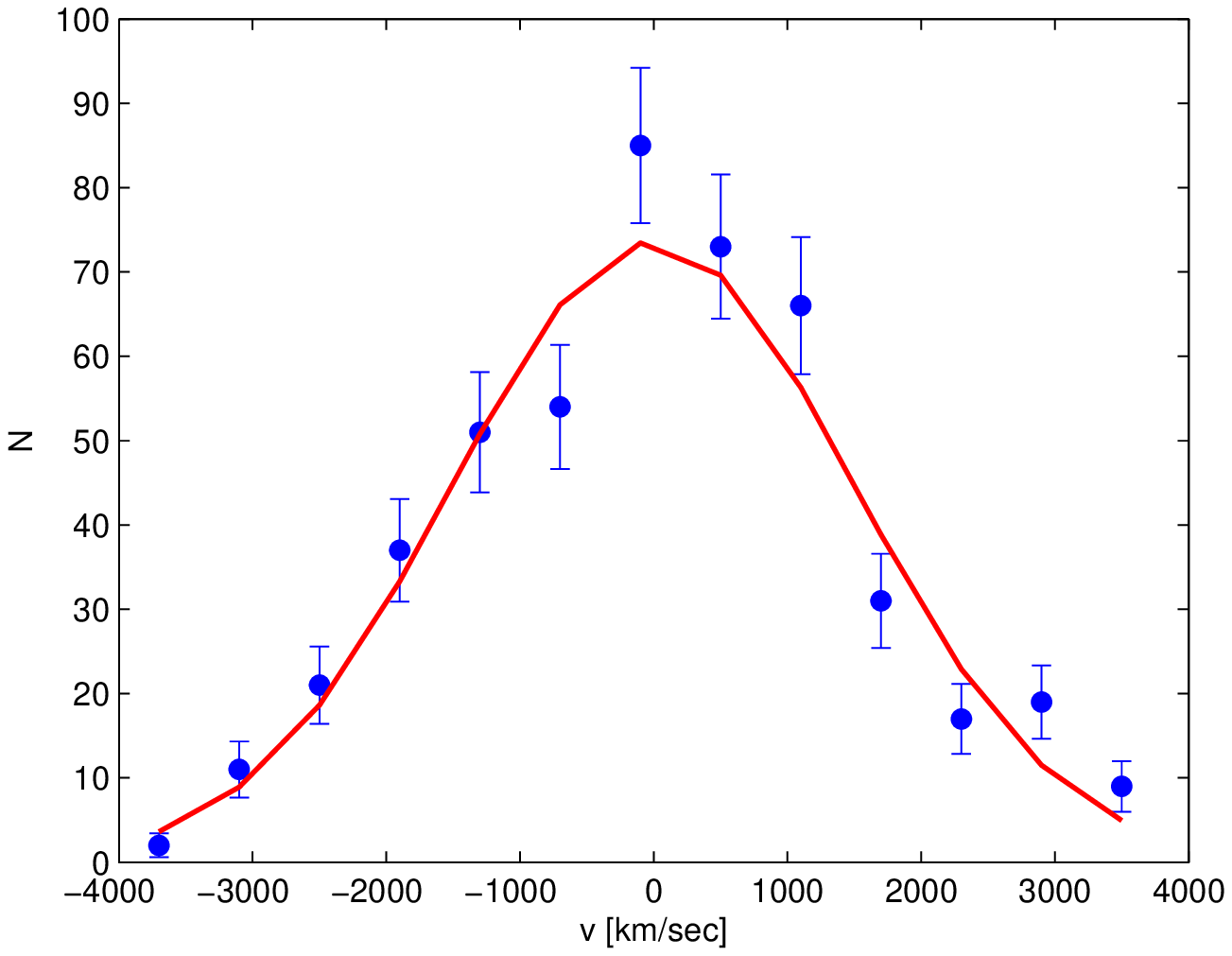, width=8cm}
\caption{Observed histogram of galaxy velocities and Gaussian fits.
Top panel: all galaxies; the fit yielded  
$\chi^2_r=2.5$ with derived parameters $z_{\rm cluster}=0.185\pm0.0003$ 
and $\sigma_p=(1.75\pm
0.08)\times 10^3$ km/sec. Bottom panel: only cluster 
members, whose identification is based on the caustics; the fit yielded 
$\chi^2_r=1.7$ with $\sigma_p=(1.40\pm 0.06)\times 10^3$ km/sec.
\label{velocity histogram}} 
\end{figure}

Having defined cluster membership based on the location of the
caustics, we can estimate the projected velocity dispersion profile
restricted to cluster members. We divided the cluster members into
velocity bins of size 600 km/sec, sufficiently large compared to the
redshift measurement uncertainty of $\sim 200$ km/sec. Different
binning gave very similar results. We fitted a Gaussian to the
velocity histogram to calculate the overall projected velocity
dispersion (figure~\ref{velocity histogram}); it provided a reasonable
fit with a reduced $\chi^2$ (hereafter $\chi^2_r$) of 1.7 per degree
of freedom. If we ignore the caustics and simply bin all galaxies
within the maximal velocity range $|v|<4000$ km/sec, the fit to a
Gaussian is significantly worse, $\chi^2_r=2.5$, with a more
asymmetric distribution and a somewhat larger velocity dispersion. 
It is important to mention that we did not use the results of the 
Gaussian fit in our analysis. We show this fit to argue in favor of the 
real physical meaning of the caustics as we have measured them, and 
demonstrate that without excluding unrelated interlopers with large 
velocities, dynamically derived quantities of the cluster may be significantly
skewed, particularly at large projected radii. Note that we fitted a Gaussian for 
simplicity, although the real distribution may be slightly non-Gaussian 
(see, e.g.,  Kazantzidis, Magorrian \& Moore 2004; Sanchis, Lokas \& Mamon 2004; 
 Diemand, Moore \& Stadel 2004; Wojtak et al.\ 2005).

We next estimated the radial profile of the projected velocity
dispersion from our sample of cluster members identified 
within the caustic boundaries. We divided the data into 10 radial bins, 
so that in each bin there were at least 47 galaxies. The radius assigned 
to each bin was the mean of the measured radii of galaxies lying within 
the bin. The projected velocity dispersion was taken to be the standard
deviation of the galaxy velocities in the bin. The error in the
projected velocity dispersion $\Delta \sigma_p(R)$ was calculated as
$\left(\Delta \sigma_p(R)\right)^2=\left( \Delta_{\rm me} \sigma_p(R)
\right)^2+\left(\Delta_{\rm sa} \sigma_p(R)\right)^2+
\left(\Delta_{\rm ce} \sigma_p (R)\right)^2$, where $\Delta_{\rm me}
\sigma_p(R)$, $\Delta_{\rm sa} \sigma_p(R)$, and $\Delta_{\rm ce}
\sigma_p(R)$ are the measurement, sample, and cluster membership
uncertainties, respectively. The measurement error is $\Delta_{\rm me}
\sigma_p(R) = 1/\sigma_p(R)\cdot dv_i/(N(R)-1)$, where $N(R)$
is the number of galaxies in the bin, and $dv_i=200$ km/sec is the
measurement error of the projected velocity of the $i$th galaxy. The
1-$\sigma$ sampling error is $\Delta_{\rm sa}
\sigma_{p\;\pm}(R)/\sigma_p(R) = [(1-2/(9\nu)\mp\sqrt{2/(9\nu)})^{-3/2}-1]$, 
where this formula is valid for $\nu\geq 30$ with $\nu=N(R)-1$ the number 
of degrees of freedom (Danese, De Zotti, \& Tullio 1980). $\Delta_{\rm ce} 
\sigma_p(R)$ is the error in the projected velocity dispersion from errors 
in estimating which galaxies are cluster members and which are
interlopers, arising from the uncertainty in the amplitude of the
caustic, $\delta A(R)$. We estimated $\Delta_{\rm ce} \sigma_p(R)$ by
calculating the difference of the projected velocity dispersion when
the cluster members are defined to lie within the caustic defined by
the amplitude $A+\delta A$ instead of $A$: $\Delta_{\rm ce}
\sigma_{p\;\pm}(R)=\left|\sigma_p(A\pm\delta A)-\sigma_p(A) \right|$.

\begin{figure}[h]
\centering
\epsfig{file=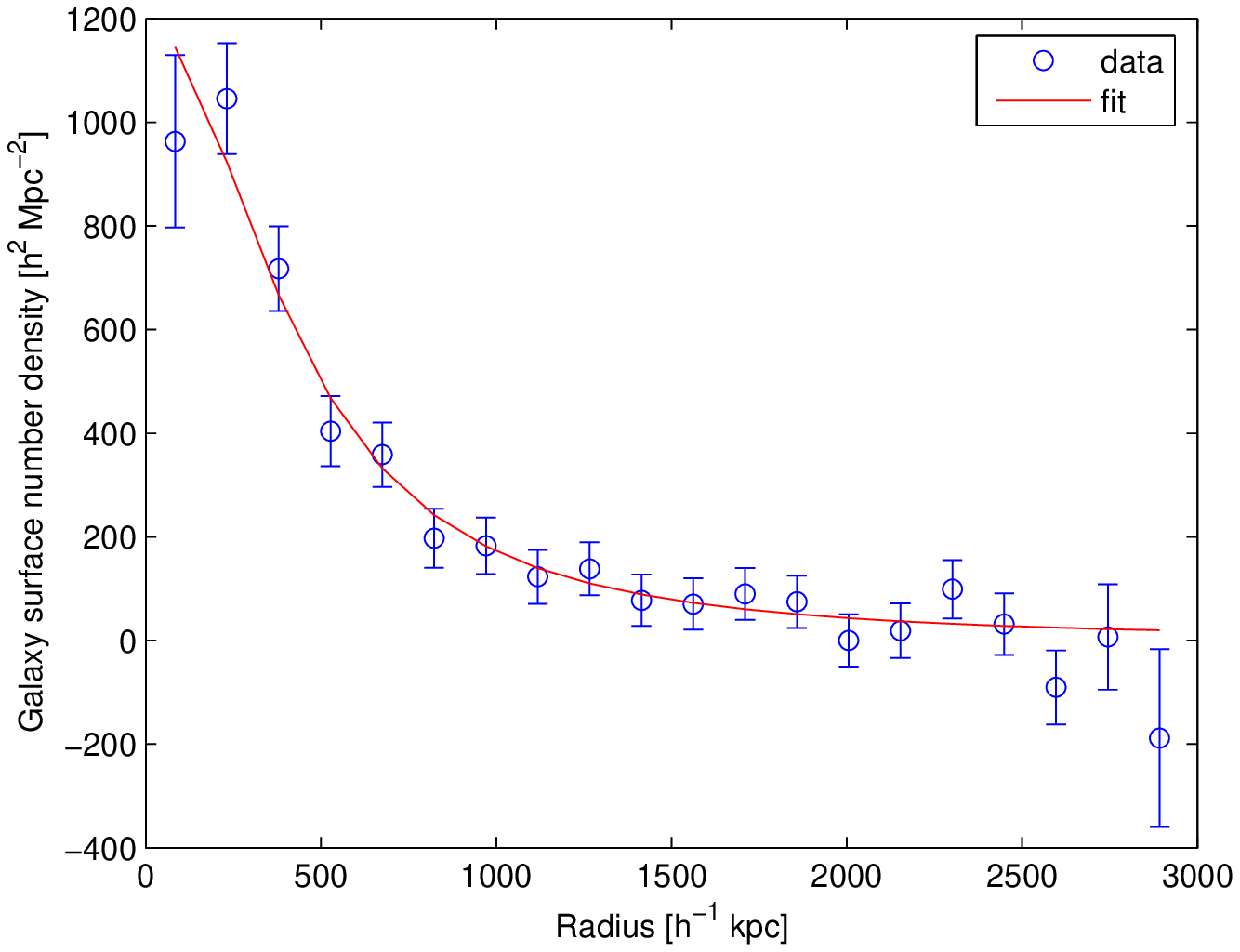, width=8cm, clip=}
\epsfig{file=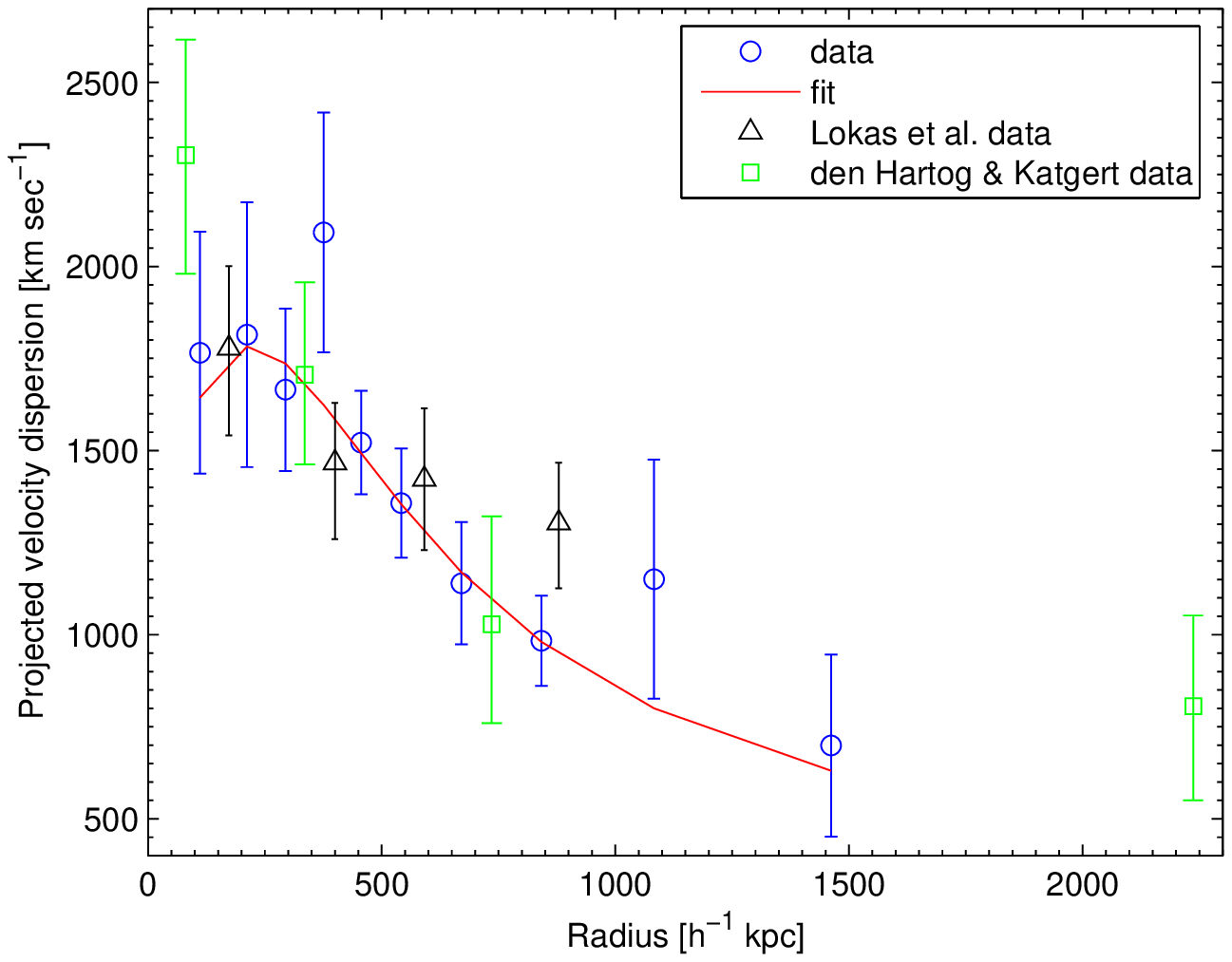, width=8cm, clip=}
\caption{Data and best fits based on analytical profiles. Top panel:
The observed profile of galaxy surface number density (blue circles) is 
compared with our best-fit model profile (red curve). Bottom panel: The 
observed profile of projected velocity dispersion (blue circles) is compared 
with our best-fit model profile (red curve). We also show the measurements 
of Lokas et al.\ (2006) (black triangles) and den Hartog \& Katgert (1996) 
(green squares). In both panels, 1-$\sigma$ measurement uncertainties 
are indicated. \label{fits to the two data sets} }
\end{figure}

There are two previous published measurements of the projected
velocity dispersion for A1689. Lokas et al.\ (2006) took all galaxies
in the NASA Extragalactic Database (NED) with redshifts
$z=0.1832\pm0.05$ and located at projected distances below 2 Mpc from
the cluster center. The redshift data for the galaxies came from
different surveys, mainly by Teague, Carter, \& Gray\ (1990), Balogh
et al.\ (2002), and Duc et al.\ (2002). While the sample of Teague,
Carter, \& Gray (1990) came from a standard magnitude-limited survey,
the selection criteria of those of Balogh et al.\ (2002), and Duc et
al.\ (2002) were aimed at star-forming galaxies, which may bias the
sample towards outer regions with more substructure. This full sample
comprised 192 galaxies, which they reduced to 130 after defining a
constant velocity cutoff of 3000 km/s, estimated visually, which is
substantially smaller than the 4000 km/s maximum amplitude of our
caustics. On this basis they claimed significant substructure in
velocity along the line of sight, but there is no evidence for such
substructure in our velocity data (Czoske 2004), nor is significant
substructure seen in the X-ray emission maps or the lensing-based mass
map (L08; Umetsu \& Broadhurst 2008). In an earlier work den Hartog \&
Katgert\ (1996) determined the radial velocity profiles of 72
clusters, including A1689. For A1689 they used the data from Teague,
Carter, \& Gray (1990) comprising only 63 cluster members. The
projected velocity profiles derived in these two papers are compared
to the one we derived in figure~\ref{fits to the two data sets}.

\section{Methodology}
\label{Methodology}

In this section we present our procedure for exploring 
the structure of the cluster using the above velocity data and the 
projected profile of the galaxy distribution derived above. In addition, 
we make use of the mass distribution that we derived in earlier work 
from a combined X-ray and lensing analysis (L08). In L08 we combined 
lensing and X-ray measurements to determine model independent profiles 
of the total mass and the gas mass. Here we add the data on the galaxy 
surface number density and the projected velocity dispersion, and relate 
them via the Jeans equation. From this we obtain the 3D galaxy number 
density profile and the velocity anisotropy profile. Our work is the first 
model-free determination of the galaxy velocity anisotropy in a cluster.

In general, we determine the best-fit values of our free parameters by
fitting the profiles of galaxy surface number density and projected
velocity dispersion simultaneously. We first proceed with simple
analytical forms for the relevant profiles, in which we fit for the
free parameters of these expressions. We then proceed to a more
flexible model-independent approach, similar to that developed in 
L08. In our model-independent approach, the free parameters are simply the 
values of the 3D profile of the galaxy number density and velocity anisotropy 
at several fixed (equally spaced) radii. The radial ranges of these free 
parameters are set according to the total span of each data set. The values 
of each quantity (i.e., the galaxy number density or velocity anisotropy) 
across the cluster are given by linearly interpolating from its values at 
the fixed radii. Thus, the method only assumes smoothness.

The galaxy number density and velocity dispersion 
are related by the Jeans equation for a steady-state spherically 
symmetric system, 
\begin{equation}
\frac{d}{dr}(n_{\rm gal}(r)\sigma_r^2(r))+\frac{2\beta(r)}{r}
n_{\rm gal}(r)\sigma_r^2(r)= -\frac{GM(\leq r)n_{\rm gal}(r)}{r^2}\ ,
\label{Jeans equation}
\end{equation} where $n_{\rm gal}(r)$ is the galaxy number density,
and $\beta(r)$ is the velocity anisotropy parameter:
\begin{equation}
\beta(r)\equiv1-\frac{\sigma_t^2(r)}{\sigma_r^2(r)}\ ,
\end{equation} where $\sigma_r(r)$ and $\sigma_t(r)$ are the radial 
and the tangential components of the velocity dispersion (Binney \&
Tremaine 1987). The measured quantities used here are the galaxy
surface number density, 
$\Sigma_{\rm gal}$,  
\begin{equation}
\Sigma_{\rm gal}(R)=2 \int^{\infty}_{R} \frac{n_{\rm gal}(r) r
    dr}{\sqrt{r^{2}-R^{2}}}\ , 
\label{eq:galaxy number surface density}
\end{equation} and the observed projected velocity dispersion, 
\begin{equation}
\sigma_p^2(R) = \frac{2}{\Sigma_{\rm gal}(R)}\int_{R}^{\infty}
\frac{n_{\rm gal}(r)\sigma_r^2(r) \left[ 1-\beta(r)
\frac{R^2}{r^2} \right] rdr}{\sqrt{r^2-R^2}}\ .
\label{projected velocity dispersion}
\end{equation}

Given a total cluster mass profile $M(r)$, for any assumed profiles of
$n_{\rm gal}$ and $\beta$ we use the Jeans equation (eq.~\ref{Jeans
equation}) to derive the profile of the radial velocity dispersion
$\sigma_r$. We then use eqs.~(\ref{eq:galaxy number surface density}) and 
(\ref{projected velocity dispersion}) to compare the model to the
two observed data sets, and find the best-fit free parameters. 
This procedure was first applied using simple analytic forms for the profiles 
of 3D galaxy number density $n_{\rm gal}$ and velocity anisotropy parameter $\beta$. 
The model-independent mass profile found in our earlier work 
(L08) was used; we also compared in some cases the effect of using its 
approximation as an NFW profile, with $c_{\rm vir}=12.2^{+0.9}_{-1}$ and 
$r_{\rm vir}=2.14^{+0.27}_{-0.29}$ h$^{-1}$ Mpc for the concentration parameter 
and virial radius, respectively (L08). For the galaxy number density profile 
we assume the $\beta$ model 
\begin{equation}
n_{\rm gal}(r)=n_{\rm gal}(0)\left[1+(r/r_s)^2\right]^{-\frac{3}{2}p}\ ,
\label{beta model}
\end{equation} 
where here we use $p$ instead of the standard $\beta$ (which we use
for the velocity anisotropy parameter); e.g., 
the King profile (King 1962) corresponds to $p=1$. N-body simulations for 
a variety of cosmologies show that the velocity anisotropy has a nearly 
universal radial profile (Cole \& Lacey 1996). We follow Carlberg et al.\ (1997)
and model it as:
\begin{equation}
\beta = (C+1)\frac{(r/r_{\rm vc})^2}{(r/r_{\rm vc})^2+1}-C\ , 
\label{beta analytic expression}
\end{equation} 
where $C$ and $r_{\rm vc}$ (a core radius of the velocity profile) 
are free parameters. The free parameter $C$ 
is needed to ensure 
the freedom for $\beta$ to take on any value in its allowed range, 
$-\infty<\beta<1$, 
including circular orbits at small radii. A similar profile is advocated by Girardi 
et al.\ (1998) for ``case A'' clusters, $\beta=\frac{r^2}{r^2+a^2}$, except 
that we do not constrain the velocities to be isotropic in the center.

When performing model-independent fits, we must extrapolate each
quantity beyond the last data point. We extrapolated the anisotropy
parameter $\beta$ to be constant, i.e., equal to its value at the last
projected velocity dispersion data point. In this way we did not
pre-constrain it to equal unity at large radii. We extrapolated the
galaxy number density $\propto r^{-3}$, consistent with a King profile
and also with the asymptotic behavior of our best-fit $\beta$ model
(which is $\propto r^{-3.18\pm0.42}$; see table~\ref{best fitted
parameters} below).

\section{The dynamical properties of the cluster galaxies}
\label{The galaxy dynamical properties}

In this section we present the results for the dynamical properties of
the galaxies, namely the galaxy number density, velocity anisotropy,
and radial velocity dispersion, all derived from our simultaneous
analysis of the galaxy surface number density and velocity dispersion
data sets. As stated in the previous section, we employed 
the Jeans equation and used the mass profile we had previously derived 
from lensing and X-ray data (with our model-independent method described in L08). 
We present here the results both from fitting simple analytical forms to 
the dynamical data and from our model-independent approach.

Using the assumed profiles, the simultaneous fit to the two data sets
was very good, as expressed by the low $\chi^2_r = 15.6/(30-5)$, where
there are 30 total data points and 5 parameters (from eqs.~\ref{beta
model} and \ref{beta analytic expression}). The contribution of each
data set within the full simultaneous 
fit to both, was $\chi^2_r({\rm number})= 12/(20-3)$ and 
$\chi^2_r({\rm velocities})= 3.6/(10-2)$. The fits are shown in 
figure~\ref{fits to the two data
sets}. The best-fit 
parameters of the analytical expressions for the galaxy number density 
and velocity anisotropy profiles are given in table
~\ref{best fitted parameters}.

\begin{table}[h]
\caption{Best-fit parameters for the galaxy 
number density and velocity anisotropy profiles \label{best fitted parameters}}
\begin{center}
\begin{tabular}{|c|c|}
\hline
   Parameter                            &  Value \\
\hline
   $n_{\rm gal}(0)$ [h$^{3}$ Mpc$^{-3}$] &  $ 1380\pm 320 $     \\
   $r_s$ [h$^{-1}$ kpc]             &  $ 455 \pm 110 $     \\
   $p$                               &  $ 1.06\pm 0.14$     \\  
   $r_{\rm vc}$ [h$^{-1}$ kpc]           &  $ 395 \pm 280 $     \\
   $C$                               &  $ 1.8\pm 2.6$     \\       
\hline
\end{tabular}
\end{center}
\end{table}
 
\begin{figure}[h]
\centering
\epsfig{file=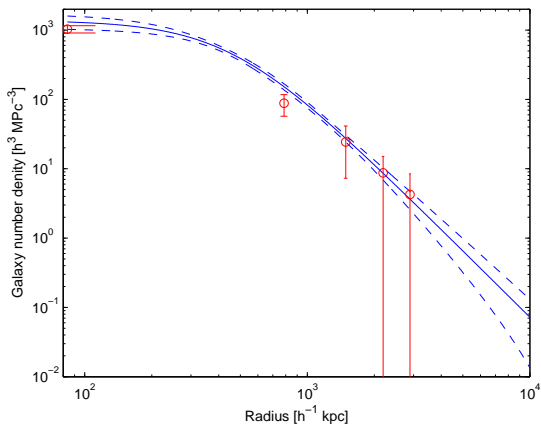, width=8cm, clip=}
\caption{Reconstructed 3D galaxy number density profile. We show the 
best-fit 
result and 1-$\sigma$ range from fitting a $\beta$ model (solid and 
dashed curves), where the best-fit model parameters are listed in 
Table~\ref{best fitted parameters}. Also shown are the values and 1-$\sigma$ 
errors as derived from the model-independent approach (red circles), where 
the free parameters were the values of the galaxy number density at $5$ equally 
(linearly) spaced radii.
 \label{3D galaxy density}
}
\end{figure}

The derived 3D profile of galaxy number density is shown in
figure~\ref{3D galaxy density}. The analytical cored profile is
consistent with our best-fit model-independent profile. We also tried 
a cuspy profile as an analytical form for the galaxy number density, 
instead of a cored profile. Specifically, we adopted a general form 
with a $1/r$ cusp, $n_{\rm gal}(r) = n_{\rm gal}(0)/[(r/r_s)(1+(r/r_s))^p]$, 
which includes NFW ($p=2$) and Hernquist (1990) ($p=3$) profiles as 
special cases. Although this profile gave an acceptable fit, 
$\chi^2_r = 26.3/(30-5)$, the fit was worse than that of the cored profile. 
The best-fit model gave $n_{gal}(0)=18.8$ and an unrealistic value for $r_s$, 
$r_s=15$ h$^{-1}$ Mpc; with this value of $r_s$ and the deduced $p$, 
$p=18.1$, the $p$-dependent term is negligible (implying that only the ratio 
$n_{gal}(0)/r_s$ is relevant).

\begin{figure}[h]
\centering
\epsfig{file=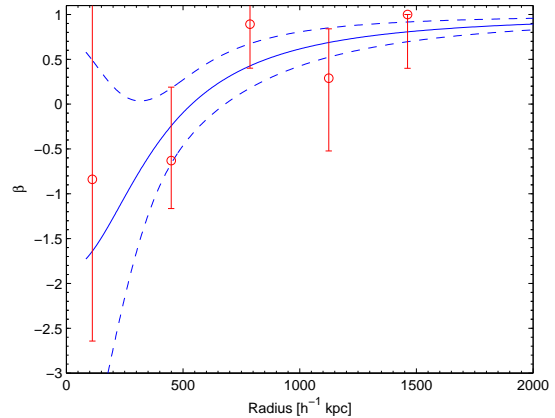, width=8cm, clip=}
\caption{Reconstructed 3D velocity anisotropy profile. We show the
best-fit result and 1-$\sigma$ range from fitting the simple profile of 
eq.~(\ref{beta analytic expression}) (solid and dashed curves), where the 
best-fit model parameters are listed in Table~\ref{best fitted parameters}. 
Also shown are the values and 1-$\sigma$ errors as derived from the model-independent 
approach, where we used 5 free parameters for the values of the galaxy 
number density and 5 (red circles) free parameters for the values of 
$\beta$ at $5$ equally spaced radii. The value of the velocity anisotropy 
was extrapolated as constant beyond the last data point.
\label{velocity anisotropy profile}
}
\end{figure}

\begin{figure}[h]
\centering
\epsfig{file=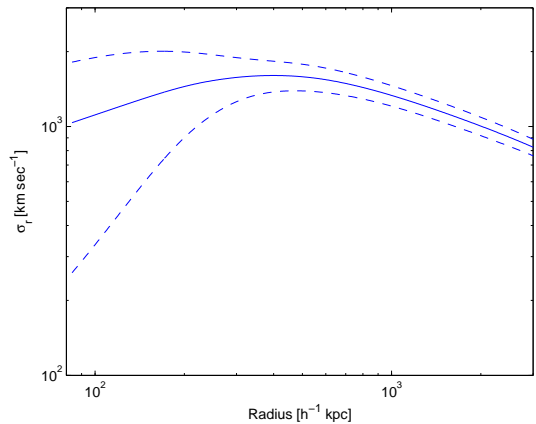, width=8cm, clip=}
\caption{Reconstructed profile of the radial velocity dispersion
$\sigma_r$. We show the best-fit 
values and 1-$\sigma$ uncertainties from fitting with the simple 
analytical expressions in eqs.~(\ref{beta model}) and (\ref{beta analytic 
expression}) (solid and dashed curves).
 \label{sigma_r profile}
}
\end{figure}

The velocity anisotropy profile $\beta$ is shown in
figure~\ref{velocity anisotropy profile}. Again, the result from
fitting a particular simple analytical form for the profile agrees
with the result of the model-independent approach which does not
assume any particular shape for the profile. 
While the uncertainties are large, the figure indicates a clear tendency 
of an increasing $\beta$ with radius. The $\sigma_r$ profile, as obtained 
from the Jeans equation (eq.~\ref{Jeans equation}), is shown in 
figure~\ref{sigma_r profile}.

The analytical expression we used for $\beta$, eq.~(\ref{beta analytic
expression}), constrains $\beta$ to equal unity at large radii,
corresponding to purely radial orbits. D99 analyzed N-body simulations
and derived values somewhat smaller than 1 at their limiting radius,
$6r_{200}$. To allow for deviation from unity we also fit the
expression $(C+a)\frac{(r/r_c)^2}{(r/r_c)^2+1}-C$, where $a$ is a free
parameter which governs the asymptotic behavior of $\beta$. The value
of $a=1$ gave the best fit, also consistent with the value obtained by
the model-independent method at the largest radial point.

\section{Deriving the mass profile}
\label{Deriving the mass profile}

In \textsection~\ref{The galaxy dynamical properties} we used the mass
profile as derived from X-ray and lensing data together with the new
data on the projected velocity dispersion and the galaxy surface
number density to derive the 3D profiles of the velocity anisotropy
and galaxy number density. Alternatively, there are at least two different 
ways to derive the total mass profile of the cluster directly from the 
data sets on galaxy dynamics. Perhaps the simplest approach 
is to use only the velocity caustics derived above, the amplitude of 
which is related to the escape velocity, which is a tracer of the 
cluster mass profile. The second approach is to use both data sets 
(of the projected velocity dispersion and galaxy surface number density) 
together with the Jeans equation to fit an NFW mass profile. Thus, we have two 
independent methods for estimating the cluster mass profile.

\subsection {Mass profile from velocity caustics}
\label{Mass profile from velocity caustics}

\begin{figure}[h]
\centering
\epsfig{file=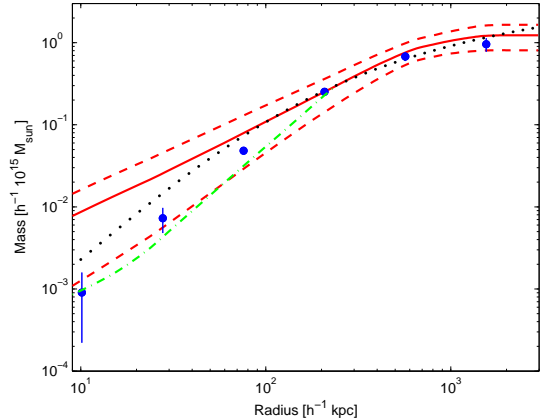, width=8cm}
\caption{Profile of total enclosed mass, as derived from the
caustics (red curve), caustics with $F_{\beta}$ corrected for $r<250$
h$^{-1}$ kpc (green dash-dotted curve), lensing and X-ray data (blue
circles, L08), and an NFW mass profile fit to the galaxy surface
number density and projected velocity dispersion data sets (black
dots). The uncertainties (error bars for the lensing and X-ray data
based points, and red dashed curves for the caustic based profile) 
are 1-$\sigma$.
\label{mass profile comparison}} 
\end{figure}

As mentioned already, D99 has shown 
that the 3D mass profile can be fairly well estimated directly from
the amplitude of the velocity caustics. In particular, the total mass
within radius $r$ is estimated as 
\begin{equation}
M(\leq r) = \frac{1}{2G}\int_0^rA^2(R)dR\ . \label{M_from_A}
\end{equation} 
This equation essentially estimates the mass profile based on the local 
escape velocity, but note that the prefactor of $0.5$ is not the result of 
an exact derivation. This prefactor comes from the more general 
suggestion by Diaferio \& Geller (1997; hereafter DG97) for 
the relation between $A$ and $M$: 
$M(\leq r) = \frac{1}{G} \int_0^r A^2(R)F_{\beta}(R)dR$ 
in terms of 
\begin{equation}
F_{\beta}(r)=-2\pi G\frac{\rho(r)r^2}
{\phi(r)} \frac{3-2\beta(r)} {1-\beta(r)}\ , \label{Fbeta}
\end{equation} 
where $\rho(r)$ is the cluster mass density and 
$\phi(r)=-\frac{GM(<r)}{r} -4\pi\int_r^{\infty}\rho(x)xdx$ 
is the gravitational potential generated by the cluster. 
D99 noted that the function $F_{\beta}(r)$ is slowly varying at large radii, 
$r \ga r_{200}/3$ (D99), when assuming the NFW profile for the cluster mass 
density and computed the anisotropy profile $\beta(r)$ from 
simulations. D99 followed DG97 and set $F_{\beta}$ to be $0.5$, finding that 
the resulting caustic method recovers the actual cluster mass within a
factor of 2 at large radii, $r \sim (0.3 - 6)\ r_{200}$, at least 
with the low concentration parameters of his simulated clusters. 
Figure~\ref{mass profile comparison} shows that for A1689 there is a 
reasonably good agreement between the mass profile derived from the 
caustics and that derived from our earlier joint lensing and X-ray 
analysis, except (for the uncorrected caustic method) in the inner 
region, although the data points even in this region are within the 
1-sigma range of uncertainty. The difference is well below a factor 
of 2 in the range $\sim (0.1- 1.5)$ h$^{-1}$ Mpc. This suggests that 
the assumption of $F_{\beta}=0.5$ is reasonable for this cluster over 
a broad radial range. Note that the NFW profile fitted to this cluster
gives a relatively high concentration parameter (see \textsection~\ref{Methodology}).

\begin{figure}[h]
\centering
\epsfig{file=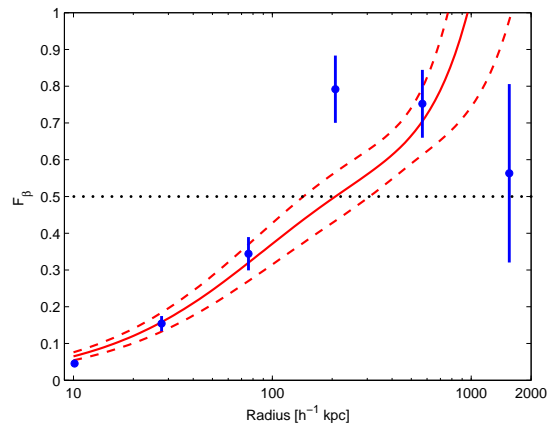, width=8cm}
\caption{Reconstructed profile of the dimensionless factor
$F_{\beta}$ given by eq.~(\ref{Fbeta}). We compare two profiles, where
we use the total mass density derived by the model-independent method in 
L08 (blue points) or the best-fit NFW profile from L08 (red curve). The 
uncertainties (blue error bars for the model-independent method and dashed 
curves for the NFW model) are 1-$\sigma$. Also shown for comparison is the 
$F_{\beta}=0.5$ (dotted horizontal) line. 
\label{F_B profile}} 
\end{figure}

At smaller radii, $r\lesssim 0.1$ h$^{-1}$ Mpc, the two mass profiles
differ by $\sim 1$-$\sigma$. This difference can arise if $F_{\beta}$
is significantly different from 0.5. As a consistency check, we can
use the velocity anisotropy profile and the mass profile of the
cluster that we derived from the lensing and X-ray analysis to
calculate the $F_{\beta}$ profile. Figure~\ref{F_B profile} shows that
at small radii ($r<0.1$ h$^{-1}$ Mpc) $F_{\beta}$ is indeed well below
0.5. In Figure~\ref{F_B profile} we compare the $F_{\beta}$ profile
derived by our model-independent method to that derived from assuming 
an NFW profile for the total mass density. The profile determined by D99 from 
simulations is somewhere in between our two measured profiles: the
simulated $F_{\beta}$ increases with radius and then flattens, but
only at large radii ($\sim \frac{1}{2} r_{\rm vir}$). Note that 
in the model-independent method, the point at the largest radius is lower 
than in the NFW case, since the density there is lower than derived using the NFW 
model due to the fact that the slope of the total density at large radii for 
A1689 is slightly steeper than for the NFW profile (B05a; L08). In 
Figure~\ref{mass profile comparison} we also plot the mass profile 
based on the caustics method when we use the more realistic $F_{\beta}$ 
value (instead of 0.5) at $r<250$ h$^{-1}$ kpc. The 
correction improves the agreement between the caustics-based mass
profile and the profile derived from the joint lensing and X-ray
analysis of L08.

The mass of the cluster derived by the caustics method is roughly 
equal to the virial mass, $M_{\rm vir} = (1.23\pm 0.42)\times 
10^{15}$ h$^{-1}$ M$_{\odot}$, since the caustics end at about 
the virial radius (see \textsection~\ref{The edge of the cluster}).

\subsection{Mass profile from the Jeans equation}
\label{mass profile from the Jeans equation}

\begin{figure}[h]
\centering
\epsfig{file=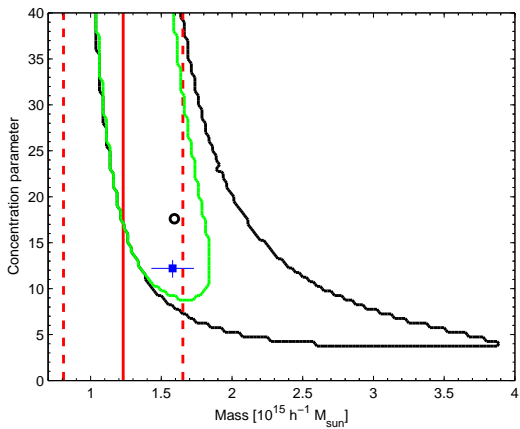, width=8cm, clip=}
\caption{Virial mass and concentration parameter, $c_{\rm vir}$, derived from NFW
fits to the cluster total mass profile. We show the results of an NFW fit 
to the galaxy dynamics using the Jeans equation (black circle for the best-fit 
values and black contour for the 1-$\sigma$ uncertainty), the mass measured at the virial 
radius from caustics (red solid and dashed lines for the best-fit value and 1-$\sigma$ 
upper limit), and the joint 1-$\sigma$ uncertainty taking both methods, using the 
Jeans equation and the caustics, under consideration (green contours). For comparison 
we show the result from the lensing and X-ray data (blue square with 1-$\sigma$ error bars; L08). 
\label{C_M_contour}
}
\end{figure}

Analysis of the lensing data suggests that A1689 has a high
concentration parameter (B05a; L08). We can independently check the
concentration parameter and the virial mass as derived solely from our
two data sets that measure the galaxy surface number density and
projected velocity dispersion. A similar attempt to do so was made by 
Lokas et al.\ (2006) but with significantly less extensive projected 
velocity data (see \textsection \ref{sec:vel}). In addition to what has 
been done in \textsection~\ref{Methodology}, we have also obtained the 
mass by implementing the following procedure: We use the galaxy number 
density and velocity anisotropy profiles (eqs.~\ref{beta model} and 
\ref{beta analytic expression}), along with an NFW profile for $M(r)$, 
which adds free parameters $c_{\rm vir}$ (the concentration parameter) 
and $M_{\rm vir}$ (the virial mass), and derive the parameters of the 
three profiles by simultaneously fitting to the galaxy number density 
and velocity data, using the Jeans equation. This method results in 
relatively weak constraints, as is clear from the rather wide 1-$\sigma$ 
contours shown in figure~\ref{C_M_contour} (black contours); the deduced 
mass and concentration are $M_{\rm vir}=1.6^{+1.1}_{-0.8}\times 10^{15}$ 
h$^{-1}$, and $c_{\rm vir}>5.8$ (with no useful upper limit).

The latter method of estimating the cluster mass, from the Jeans
equation with assumed profiles, is less precise than the other
methods, yielding a partial degeneracy between $c_{\rm vir}$ and
$M_{\rm vir}$ as shown in figure~\ref{C_M_contour}. We thus expect
this constraint to be weak. The observed galaxy number density
essentially determines $n_{\rm gal}(r)$ through eq.~(\ref{eq:galaxy
number surface density}), and the observed projected velocity
dispersion determines a degenerate combination of $\sigma_r^2$ and
$\beta$ at each radius through eq.~(\ref{projected velocity
dispersion}). This yields a degeneracy where for any assumed
$\beta(r)$, the Jeans equation yields an $M(r)$ that is consistent
with the galaxy dynamical data. In the actual fitting though, this
degeneracy is partially broken by the strict analytical forms assumed
for the various input profiles. Figure~\ref{mass profile comparison}
shows that the best-fit NFW mass profile from the Jeans equation is in
good agreement with the mass profiles from the other methods.

To get more useful constraints on $c_{\rm vir}$ and $M_{\rm vir}$, we
combine the two dynamical methods, i.e., the caustics and the Jeans
equation. From the caustics method we take only the constraint on
$M_{\rm vir}$, since the mass profile at small radii is uncertain in
this method due to the breakdown of the $F_{\beta}=0.5$ assumption
(see discussion in the previous subsection). Figure~\ref{C_M_contour}
shows that the combined constraints (green contours) are 
stronger and in good agreement with the values derived from the lensing and
X-ray data by L08 (who assumed in this particular analysis an NFW
profile for the total mass density and a double beta model for the gas
mass density profile). In particular, the combined dynamical methods
yield 1-$\sigma$ limits of $M_{\rm vir}=(1.3\pm 0.4) \times 
10^{15}$ h$^{-1}$ M$_{\odot}$ and $c_{\rm vir}>13.4$.

\section{The edge of the cluster}
\label{The edge of the cluster}

In this section we summarize various ways of defining a limiting
radius for A1689. One way to define the edge of the cluster is to use
the observed galaxy number density profile. The fits that we have used
in eqs.~(\ref{Sigma_tot}) and (\ref{beta model}) do not have a sharp
cutoff in the number density of galaxies. We thus define an edge as
the radius where we can no longer detect cluster members above the
contribution of the (non-cluster) background galaxy level. For A1689
this point is visible in figure~\ref{galaxy surface number density}.
Specifically, the cluster radial edge was estimated to be where
$\Sigma_{\rm gal}/C_{\rm bg}>\Delta C_{\rm bg}/C_{\rm bg}$ and 
$\Sigma_{\rm gal}/C_{\rm bg}<\Delta C_{\rm bg}/C_{\rm bg}$ including 
uncertainties in $\Sigma_{\rm gal}/C_{\rm bg}$, 
yielding a limiting
radius of $2.1_{-0.7}^{+0.8}$ h$^{-1}$ Mpc, where the 1-$\sigma$
uncertainties account also for the errors in the various fitting parameters
in eq.~(\ref{Sigma_tot}).

Independently, the velocity caustic fits to the projected velocity
dispersion data shown in figure~\ref{velocity space diagram} yield a
very similar value for the limiting radius, $2.12\pm0.07$ h$^{-1}$
Mpc, where the error includes an estimate of the effect of Poisson
noise in the observed number of galaxies. We caution that in
simulations the caustics often flatten but do not reach zero at the
virial radius; also, the shape of the caustics is somewhat dependent
on the particular line of sight (D99). However, the caustics are
generally more cleanly defined in data on real clusters than in N-body
simulations (Rines et al.\ 2003).

Both of these methods yield a cluster edge limiting radius of $\sim 2$ 
h$^{-1}$ Mpc. A similar value was also independently derived from our 
lensing and X-ray analysis, which depends mostly on the projected DM 
distribution. We found the virial radius to be $2.14^{+0.27}_{-0.29}$ 
h$^{-1}$ Mpc (L08). We conclude that all these different data sets 
agree reasonably well both in terms of the virial mass and mass profile 
and in the values they yield for the cluster's limiting radius.

The various ways of defining a limiting radius for A1689 are
consistent with each other. This suggests that there is no major
infall of DM and galaxies. Significant ongoing infall would add to the
projected profiles of mass and galaxy number and also affect the
dynamical measurements, likely making the cluster edge less apparent.

\section {Discussion}
\label{Discussion}

In this paper we have continued our exploration of A1689 making use of
many high quality datasets available for this cluster. This work
builds upon our earlier work on this cluster (L08), where we developed
a comprehensive joint analysis of high quality strong lensing
(HST/ACS), weak lensing (Subaru), and X-ray (Chandra) measurements,
from which we tested the consistency of X-ray and lensing data in a
model independent way and derived an improved mass profile for
A1689. In this paper we have incorporated two other high quality data
sets, the galaxy surface number density measured from deep, wide-field
imaging with Subaru/Suprime-Cam and a large spectroscopic study of the
internal galaxy dynamics measured using VLT/VIMOS. While the lensing
and X-ray data gave us the information on the DM and gas content of
the cluster, the two new data sets added here provide direct
information on the galaxy distribution and motions, leading to new
determinations of the DM distribution of the cluster.

The 3D galaxy number density profile derived from our combined 
analysis of the above datasets is more consistent with 
a cored profile, rather than a cuspy profile such as the NFW profile, 
which seems to fit well the DM density distribution. This is in agreement 
with Adami et al.\ (1998) who examined a sample of 62 clusters and found 
that the majority are better fit with a core than a cuspy profile, though 
for individual clusters the preference for a cored profile is rarely significant at
the 90\% confidence level. The galaxy distribution also resembles a
King (1962) profile and falls off as $r^{-3.18\pm 0.42}$, exceeding
the slope of $r^{-2.4\pm0.2}$ suggested by Bahcall \& Lubin (1994) in
order to explain the "$\beta$-discrepancy". This asymptotic behavior
of the cluster galaxy profile at large $r$ is in fact very similar to
our total matter profile (dominated by DM), which we have shown can be
well fit by an NFW profile where the asymptotic behavior is $r^{-3}$.

A principal finding of our work is the first direct determination of the velocity 
anisotropy profile for a galaxy cluster. This followed from an application 
of the Jeans equation, using as input the observed projected velocity 
dispersion profile, the observed projected galaxy distribution, and 
our independently determined mass profile, allowing us to solve for the 
3D velocity anisotropy as a function of radius. The resulting anisotropy 
profile is well fit by the expression in equation~(\ref{beta analytic expression}) 
proposed by Carlberg et al.\ (1997) on the basis of N-body simulations. The 
simulations covered a wide range of cosmologies and showed that the 
radial dependence of the velocity anisotropy $\beta$ has a nearly universal 
form (Cole \& Lacey 1996; Carlberg et al.\ 1997), with a characteristic 
radial dependence. This dependence is also observed in A1689 - mainly radial 
motion is deduced at large radius, tending towards isotropic (or possibly 
tangential) motion within the central region $r \la 500$ h$^{-1}$ kpc. 
This presumably is a manifestation of the 
overall formation and growth of clusters, with initial collapse and 
virialization of the central region, and continued growth of the cluster 
mass through accretion, a two-stage process for which there seems to be 
some evidence also from the DM entropy distribution (e.g., Lapi \& 
Cavaliere 2008).

Additionally, we used new extensive measurements of galaxy positions 
and velocities to determine the cluster mass profile. This was done in 
two independent ways; first, using the velocity data alone, we identified 
clearly apparent velocity caustics using the method of D99. The derived 
amplitude of the velocity caustics was interpreted as the local escape 
velocity, from which the mass profile was determined using eq.~(\ref{M_from_A}). 
Secondly, we followed the traditional approach of using the Jeans equation 
(eq.~\ref{Jeans equation}), incorporating both the galaxy surface number 
density and the projected velocity data, and adopting the above velocity 
anisotropy profile. These two different mass estimates are in good agreement 
with the profile derived from our earlier lensing and X-ray analysis (L08), 
as shown in figure~\ref{mass profile comparison}. 

In estimating the caustic-based mass profile we adopted the previously 
suggested value $F_{\beta}=0.5$. We were able to separately check this 
assumption using the velocity anisotropy profile obtained as described above. 
Our previously determined mass profile made it possible to deduce $\beta$; 
the comparison of the resulting mass profile with that from our previous 
lensing/X-ray analysis is then essentially a consistency check on the 
general validity of the caustic method. We found that at large radii, 
$r \gtrsim 100$ h$^{-1}$ kpc, the factor $F_{\beta}$ varies slowly with 
radius and stays within $\sim 50\%$ of the value of 0.5 (see figure~\ref
{F_B profile}), so that the relatively simple caustic method yields the 
mass profile accurately except at the center. Note that D99 derived the 
mean value of $F_{\beta}=0.5$ from simulated CDM halos, which typically 
have concentration parameter $\la 7$, well below that of A1689 
($c_{\rm vir}=12.2^{+0.9}_{-1}$; L08). Thus, our results suggest that
the caustic-based mass estimation is applicable also for high concentration 
clusters. This shows that dynamical analysis may be improved upon by 
combining the traditional method based on the Jeans equation with the 
additional insight gained from the caustics (in relaxed clusters).

The virial mass derived from the caustic method, $M_{\rm vir} = 
(1.23\pm 0.42)\times 10^{15}$ h$^{-1}$ M$_{\odot}$, and from the Jeans
equation, 
$M_{\rm vir}=1.6^{+1.1}_{-0.8} \times 10^{15}$ h$^{-1}$ M$_{\odot}$. 
The two combined dynamical methods gave 
$M_{\rm vir}=(1.3\pm 0.4) \times 10^{15}$ h$^{-1}$ 
M$_{\odot}$, as compared with the value obtained in our joint 
lensing/X-ray analysis (L08), 
$M_{\rm vir}=(1.58\pm 0.15) \times 10^{15}$ h$^{-1}$ M$_{\odot}$. 
These mass estimates are consistent, and in agreement also 
with the result of Umetsu \& Broadhurst (2008), who combined strong
lensing, weak lensing distortion and magnification data in a 2D
analysis (without assuming axial symmetry), and derived 
$M_{\rm vir} =1.5^{+0.6}_{-0.3} \times 10^{15}$ h$^{-1}$ M$_{\odot}$ 
(where this $1-\sigma$ error includes both statistical and systematic
uncertainties).

A novel aspect of our work is an estimation of the limiting radius of
A1689 in several different ways. All our data sets independently
indicate that this cluster has a relatively well defined edge radius
of $\sim 2.1$ h$^{-1}$ Mpc. We see this in terms of the galaxy
distribution which is flat beyond this radius, indicating that the
galaxy number density reached the background level, from the velocity
caustics which attain (practically) zero velocity at this radius, and
also from the virial radius derived from the best-fit NFW
profile. This interesting agreement should be checked in other massive
clusters. In principle, we may expect a clear signature of the virial
radius in the absence of significant ongoing infall of galaxies and
mass onto the cluster.

Massive clusters ($M\sim10^{15}$ h$^{-1}$ $M_{\odot}$) 
are theoretically expected to have relatively low mass concentrations, 
with $c_{\rm vir}(z=0)\sim$ 5, as determined from cosmological simulations 
of the standard $\Lambda$CDM (Neto et al.\ 2007; Hennawi et al.\ 2007; 
Duffy et al.\ 2008). However, significantly higher values, $c_{\rm vir}\sim 10-15$, 
were deduced from detailed weak and strong lensing measurements of several well-studied 
massive clusters (Kneib et al.\ 2003; Gavazzi et al.\ 2003; B05a; Kling et al.\ 2005; 
Limousin et al.\ 2007; L08; Umetsu \& Broadhurst 2008; Broadhurst et al.\
2008). This discrepancy can also be expressed in terms of the Einstein
radius, which provides a model-independent measure of the central mass
within clusters, and can be compared with the total (virial) mass to
get a measure of the degree of concentration. The observed Einstein
radii of many massive clusters lie in the range $20^{\prime\prime}-50^{\prime\prime}$, 
around twice the expected range after allowance for lensing and projection biases 
(Broadhurst \&Barkana 2008). In addition, Sadeh \& Rephaeli (2008) showed that even
considering the full probability distribution function of halo
formation times does not fully remove the discrepancy.

While we have not measured the concentration parameter of A1689 as accurately 
as from the lensing data, we have confirmed its high value. The two 
combined dynamical methods yield $c_{\rm vir}>13.4$ (at 1-$\sigma$ confidence), 
consistent with $c_{\rm vir} = 12.2^{-1}_{+0.9}$ derived in our joint 
lensing and X-ray analysis (L08), and with $c_{\rm vir} = 12.7 \pm 3$ 
from Umetsu \& Broadhurst (2008). Clearly, this comparison should be 
examined for other well-studied clusters, but at least in A1689 this 
independent dynamical measurement, which is less susceptible to projection 
bias than lensing, is in good agreement with the lensing results, 
and can be seen (at least partly) as evidence for the validity of 
(the assumption of) spherical symmetry.

The effect of triaxiality may significantly lower the observationally 
deduced concentration parameter (Oguri et al.\ 2005; Gavazzi 2005).
An independent analysis of A1689 by Corless, King, \& Clowe (2008) does not find this to be 
significant, $c_{vir}=12.2\pm6.7$, in good agreement with our earlier 
analysis, where the sizable uncertainty reflects the unknown triaxiality and the
relatively shallow imaging used. An important step in
empirically examining the effect of triaxiality and geometric
projection bias is to see whether there is any systematic difference
between the distribution of mass profiles derived for clusters
selected by different methods. In a recent analysis of a cluster
sample selected using weak lensing, Hamana et al.\ (2008) -- who used
the sample of Miyazaki et al.\ (2002) -- did not find evidence for
selection bias in mass estimates between clusters selected by weak
lensing as compared with other selection methods, such as optically or
by X-ray emission. Also, Duffy et al.\ (2008) found that the derived
concentrations for X-ray selected samples taken from the literature
lie well above the expected values based on cosmological N-body
simulations with the most recent cosmological parameters.

At least some of the high deduced values of $c_{vir}$ are expected 
in early dark energy (EDE) models. In these models there is a non-
negligible dark energy component even at high redshift, in contrast 
with the standard model; consequently, clusters form earlier and are 
more abundant than predicted in $\Lambda$CDM (e.g, Sadeh \& Rephaeli 
2008). In recent work using N-body simulations it was found that halos 
in an EDE model show higher concentrations at a given halo mass (Grossi 
\& Springel 2008). Interestingly, in contrast with expectations from 
semi-analytic approaches, recent simulations show that EDE does not 
significantly affect the statistics of non-linear structures, at least 
at low redshifts (Francis, Lewis, \& Linder 2008; Grossi \& Springel
2008). In addition, the measured X-ray background cannot rule out EDE
models if their normalization is defined by the halo density at $z=0$
(Lemze, Sadeh, \& Rephaeli 2009). We intend to explore further the issue 
of values of $c_{vir}$ by extending our dynamical work to additional 
well-studied clusters for which we have obtained precisely measured 
lensing-based mass profiles.

\section*{ACKNOWLEDGMENTS}

We thank Oliver Czoske for providing the redshift survey information for 
A1689 and for many helpful discussions. We also acknowledge discussions 
with Sharon Sadeh, Antonaldo Diaferio, Masataka Fukugita, Hagai Netzer,
and the referee for useful comments. We wish to thank the referee 
for a thorough reading of  a previous version of the paper, and for several 
useful suggestions. This paper was partly written while DL and TB were 
visitors of the Institute for the Physics and Mathematics of the Universe, 
University of Tokyo, which we thank for its generous hospitality. RB is 
grateful for support from the ICRR in Tokyo, Japan, the Moore Distinguished 
Scholar program at Caltech, and the John Simon Guggenheim Memorial Foundation.
KU is partially supported by the National Science Council of Taiwan under the
grant NSC97-2112-M-001-020-MY3.

\end{document}